\documentclass[a4paper,11pt]{article}

\usepackage{pos}

\title{Status and results of the prototype LST of CTA}
 \ShortTitle{LST Status}

\author*[a,b]{D. Mazin} 

\affiliation[a]{Institute for Cosmic Ray Research, University of Tokyo, 5-1-5, Kashiwa-no-ha, Kashiwa, Chiba 277-8582, Japan.}
\affiliation[b]{Max-Planck-Institut f{\"u}r Physik, 80805 M{\"u}nchen, Germany.}

\forColl{CTA LST} 

\emailAdd{mazin@icrr.u-tokyo.ac.jp}

\abstract{The Large-Sized Telescopes (LSTs) of Cherenkov Telescope Array (CTA)
are designed for gamma-ray studies focusing on low energy threshold,
high flux sensitivity, rapid telescope repositioning speed and a large field of view.
Once the CTA array is complete, the LSTs will be dominating the CTA performance between 20 GeV and 150 GeV.
During most of the CTA Observatory construction phase, however, the LSTs will be dominating the array performance until several TeVs.
In this presentation we report on the status of the LST-1 telescope inaugurated in La Palma, Canary islands, Spain in 2018.
We show the progress of the telescope commissioning, compare the expectations with the achieved performance, and give a glance of the first physics results.}

\FullConference{37$^{\rm{th}}$ International Cosmic Ray Conference (ICRC 2021)\\
		July 12th -- 23rd, 2021\\
		Online -- Berlin, Germany}

\begin{document}
\maketitle

\section{Introduction}
\label{sec:intro}

\begin{figure}
\begin{center}
\includegraphics[width=\textwidth]{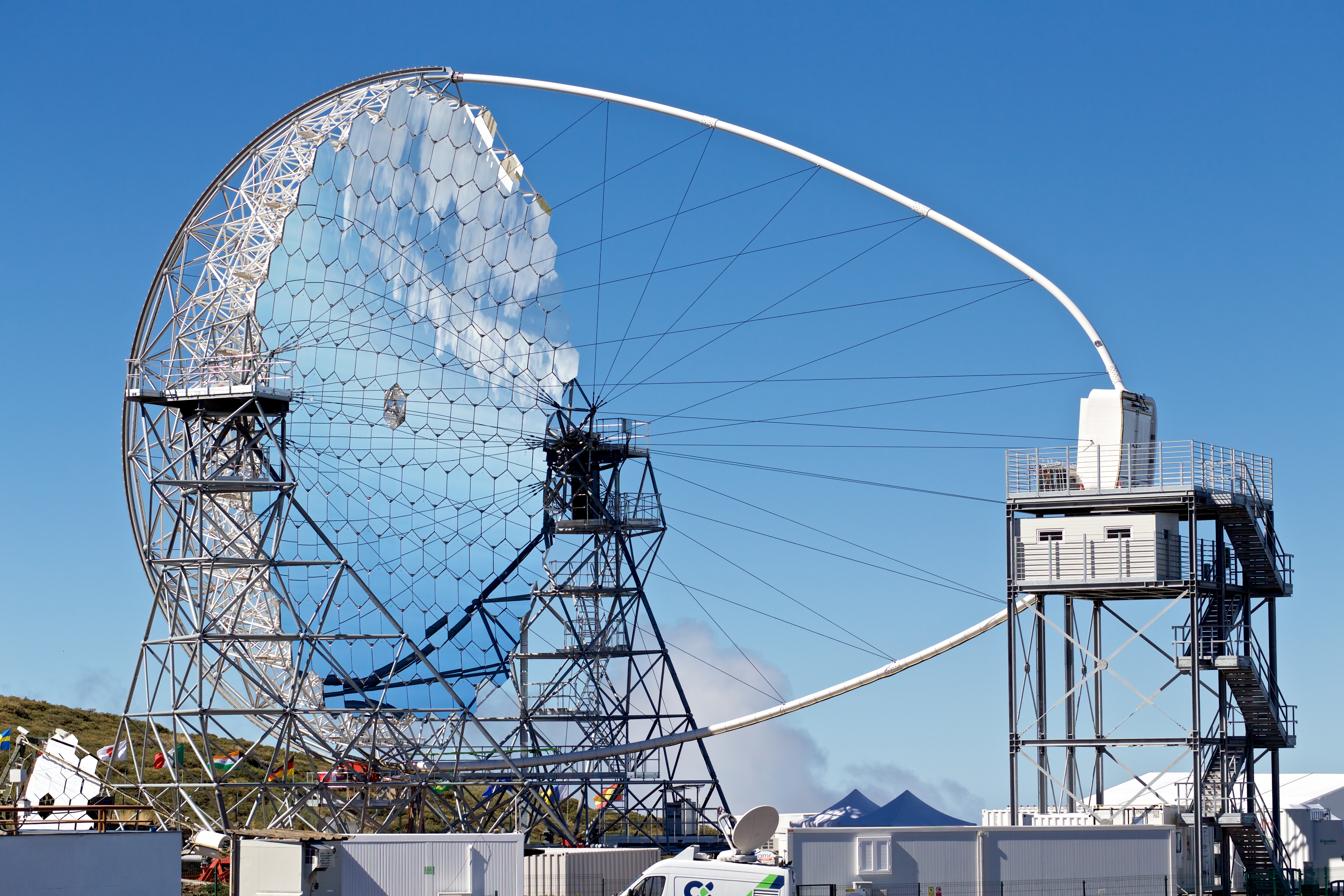}
\vspace{-0.3cm}
\caption{Photograph of the LST1 telescope in La Palma during the inauguration in October 2018. Courtesy of Akira Okumura.}
\label{fig:LST1}
\end{center}
\end{figure}

Cherenkov Telescope Array (CTA) is the next generation world wide observatory built on two sites: La Palma, Spain in the northern hemisphere and Paranal, Chile in the southern hemisphere. CTA will consist of arrays of Imaging Atmospheric Cherenkov Telescopes (IACTs) of different sizes, optimised for different energy ranges within the Very-High Energy (VHE) band. The largest telescopes of the array will be the LSTs, see Figure~\ref{fig:LST1}, with a focal length of 28\,m and a parabolic mirror dish of 23\,m diameter. LSTs target the lowest energies accessible to CTA, down to a threshold energy of $\simeq$20\,GeV. For general CTA status and plans see \citep{general_cta}.  

LST-1, the first LST, is located in the CTA North site, at a height of 2200 m a.s.l. on the Roque de los Muchachos Observatory (ORM) on the Canary Island of La Palma, Spain (28$^\circ$N, 18$^\circ$W). The first LST1 sky observations took place in December 2018, while regular LST1 observations began in November 2019. 

This paper is structured as follows. In Section~\ref{sec:project} we describe the LST project.
The commissioning milestones of LST1 are outlined in Section~\ref{sec:commissioning}. In Section~\ref{sec:performance} we describe the achieved telescope performance and compare it with the expectations. The status of the LST2-4 procurement and construction 
is given in Section~\ref{sec:lst24}, while we conclude in Section~\ref{sec:conclusion}. 


\section{LST project}
\label{sec:project}



The LST project for CTA has been developed since 2006 under leadership of MPI for Physics, Munich,
ICRR, University of Tokyo, CIEMAT Madrid, IFAE Barcelona, INFN Pisa, INFN Padova, and LAPP, Annecy. Over the years the project grew
and currently consists of about 30 institutes from 12 countries: 
Brazil, Bulgaria, Czech Republic, Croatia, France, Germany, India, Italy, Japan, Poland, Spain, and Switzerland with more than 300 scientists and engineers involved. 
The telescope design is focused on two major aspects: (1) the fast repositioning of the telescope for reaction to transient phenomena,
and (2) high sensitivity to gamma rays at low edge of the VHE range. In fact, in the CTA design \citep{cta_main} LSTs cover the low energy end, dominating the CTA performance between 20\,GeV and $\sim$200\,GeV.

In order to achieve the desired performance, LST has a light-weight structure, parabolic dish, and a power system 
based on 2x300\,kW VYCON FlyWheels to allow for the fast rotation to catch transient events like Gamma-Ray Bursts.
LST is an alt-azimuth telescope, the parabolic reflective surface is supported by a tubular structure made of carbon fiber and steel tubes. The total moving weight of the telescope (excluding the rail) is around 100 tons.
A reflective surface of $\sim$400\,m$^{2}$ is made of 198 mirror segments, which can be aligned independently 
using Active Mirror Control (AMC) system, consisting of two actuators each to compensate for small dish deformations at different zenith angles. The optical dish collects and focuses the Cherenkov photons into the Camera, where 1855 photo-sensors convert the light in electrical signals that can be processed by dedicated electronics. 
The Camera has a field of view of about 4.5$^{\circ}$ and has been designed for maximum compactness and lowest weight, cost and power consumption. Each pixel incorporates a light guide, a photo-sensor and the corresponding readout electronics. The electronics is based on the DRS (Domino Ring Sampler) chip \citep{drs4}, which, e.g., is currently used in the MAGIC experiment \citep{magic}. The Camera trigger strategy is based on the shower topology and the temporal evolution of the Cherenkov signal produced in the Camera. The analogue signals from the photosensors are conditioned and processed by dedicated algorithms that look for extremely short (few ns) light flashes. Furthermore, the LST Cameras have a hardware trigger connection in order to form an on-line coincidence trigger among the telescopes, which allows to suppress accidental triggers by up to a factor of $\approx$100.

The LST1 was built on La Palma between 2016 and 2018 as a CTA prototype telescope and was inaugurated in October 2018. Since then LST1 is in commissioning phase performed by the LST project members. 

\section{LST1 commissioning}
\label{sec:commissioning}



\begin{figure}
\begin{center}
\includegraphics[width=0.82\textwidth]{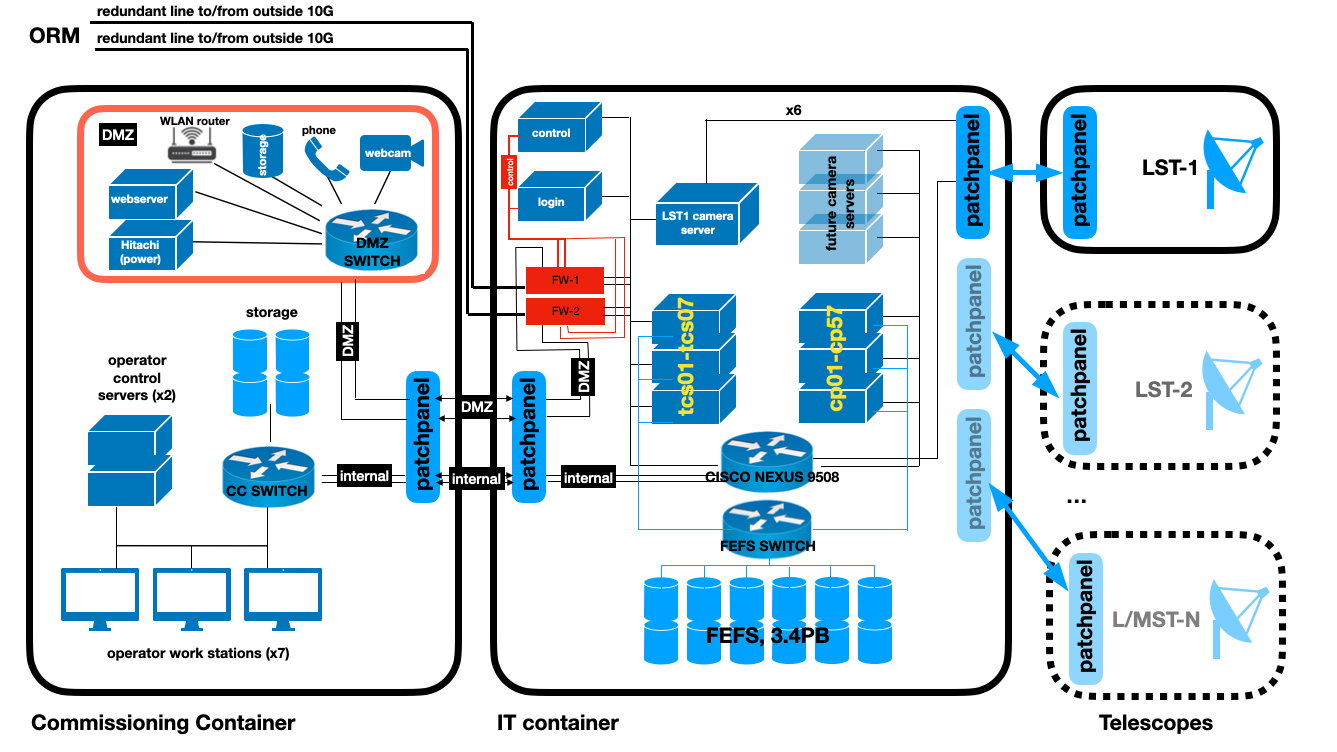}
\vspace{-0.3cm}
\caption{Schematics of the data network installed at the LST1 site.}
\label{fig:network}
\end{center}
\end{figure}

\begin{figure}
\begin{center}
\includegraphics[width=0.64\textwidth]{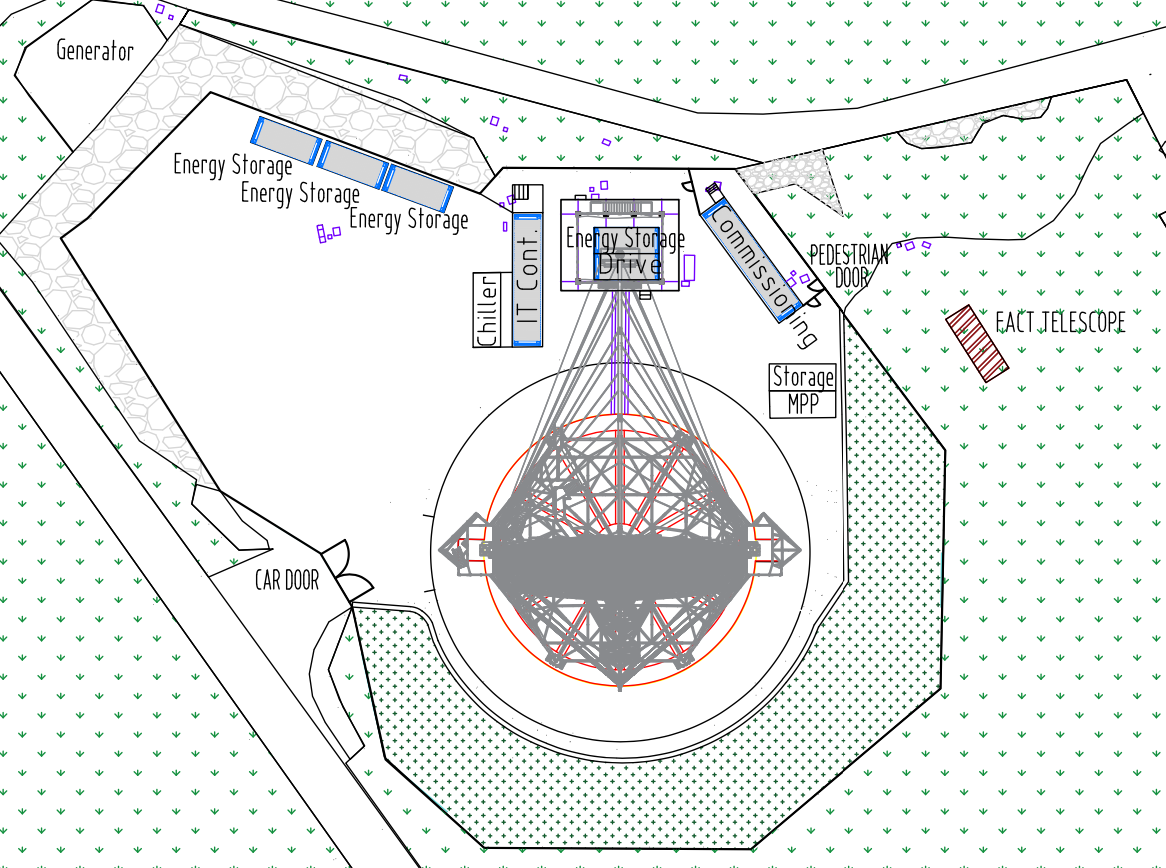}
\vspace{-0.3cm}
\caption{Top view on the site of LST1 in La Palma.}
\label{fig:infra}
\end{center}
\end{figure}

The LST1 has been built at the ORM and is profiting from the infrastructure of the existing observatory: roads, electricity, data connection, a residence for workers, and maintenance facilities. 
On the other hand, at the time of the LST1 construction no CTA array facilities, such as dedicated onsite data center, power lines, or operation building existed.
Therefore, members of the LST project invested into the missing items. Most importantly, a containerized data center, dubbed IT container, with 2000 cores, a disk storage space of 3.4\,PB and a SLURM batch job system has been installed and put into operation by the University of Tokyo. The data network, which can serve in its current configuration 4~LSTs and up to 9~Medium-Sized Telescopes (MSTs) is schematically shown in Figure~\ref{fig:network}. Moreover, a 40~foot commissioning container with 4 working spaces has been installed onsite to allow for telescope commissioning and technical operations. 
In addition, two 20 foot storage containers for spare material and tools have been temporarily placed in the LST1 site until the CTA operation building becomes ready. 
The overview of the infrastructure at the LST1 site is given in Figure~\ref{fig:infra}. Three energy storage containers are seen on the left side, ready for the LST2-4 telescopes. It is planned that the hardware of the IT center as well as the functionality of the commissioning container will be moved in the CTA operation building. 

The LST members have set up the telescope control system, the data acquisition, monitoring auxiliary information and data processing in close cooperation with CTA observatory staff in order to optimize resources and to avoid duplication of work. The low level control software is mainly based on OPCUA while higher level control is based on the ACS layer developed for ALMA radio observatory. LST members also commissioned an automatic alert system to react to transient phenomena seen in other wavelengths \cite{transient}.
The onsite IT center allows LST members to run data processing pipelines and develop new analysis methods using LST1 data and Monte Carlo simulations. LST data are also copied to PIC in Barcelona and CNAF in Bologna as backup and for possible later reprocessing.

The LST commissioning has the following goals:
\begin{itemize}
    \item Ensure compliance of the telescope elements and operations with the safety concepts and regulations. 
    \item Verify CTA scientific requirements and specifications. The LST1 is a prototype telescope and several requirements need to be verified using this telescope before producing further LSTs. In particular, the telescope bending and pointing accuracy, the optical point spread function and the fast movement are items which cannot be easily prototyped in a laboratory.
    \item Fine tune the telescope scientific performance towards a robust operation at a lowest energy threshold keeping the telescope availability at larger than 95\% of the observation time. 
\end{itemize}
After the initial phase in which the individual subsystems have been developed and debugged for a stable operation, the focus in the last 12 months is on automatising and on simplification of the control routines.

The LST commissioning is split in three main phases.
\begin{itemize}
    \item In the first phase, the commissioning is performed by the experts only, focusing on safety aspects and basic functionality. Most of the work is done during the day.
    \item In the second phase, some limited night observations are added and two crews work together: experts during the day and operators during the first half of the night. The teams overlap in the late afternoon to ensure the knowledge transfer.
    \item In the third phase, experts leave the site and provide help from remote in case of problems. The telescope operation is performed during dark nights and nights with a partial moon light. The onsite (or offsite) shift crew rotates according to the Spanish work regulations. 
\end{itemize}
The first phase lasted from Autumn 2018 until Summer 2019. The second phase started in Spring-Summer 2019, and the third commissioning phase started in January 2020.

\paragraph{COVID-19 and remote operations}

Due to the COVID-19 pandemic, the commissioning has been halted between March and June 2020. 
During that time the working places inside the commissioning container have been adapted to comply with the new sanitary regulations.
On top of that, during 2020 several remote operation locations have been installed: in Tokyo, Kyoto (Japan), Rijeka (Croatia), Annecy (France), and Torino (Italy). The onsite operation staff has been reduced to the absolute minimum to ensure the telescope safety in case of technical troubles and the operation and commissioning load have been shifted to remote operations. This concept proved to work very well thanks to a stable internet connection to La Palma and usage of novel remote desktop tools such as NoMachine NX allowing several simultaneous remote connections to the same desktop as well as good audio and video transmission quality.

\section{LST1 lessons learned and achieved performance}
\label{sec:performance}


\begin{figure}
\begin{center}
\includegraphics[width=0.49\textwidth]{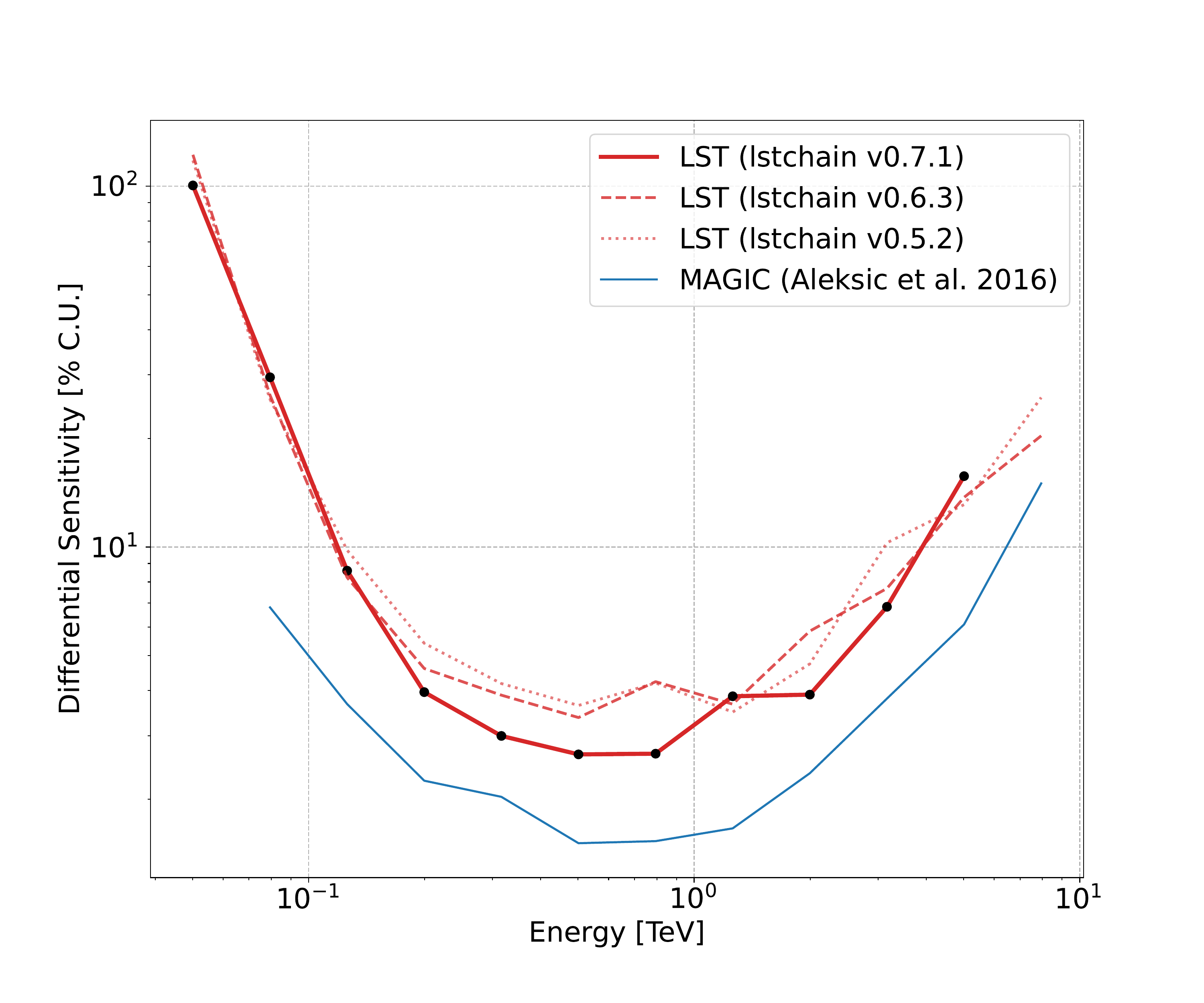}
\includegraphics[width=0.50\textwidth]{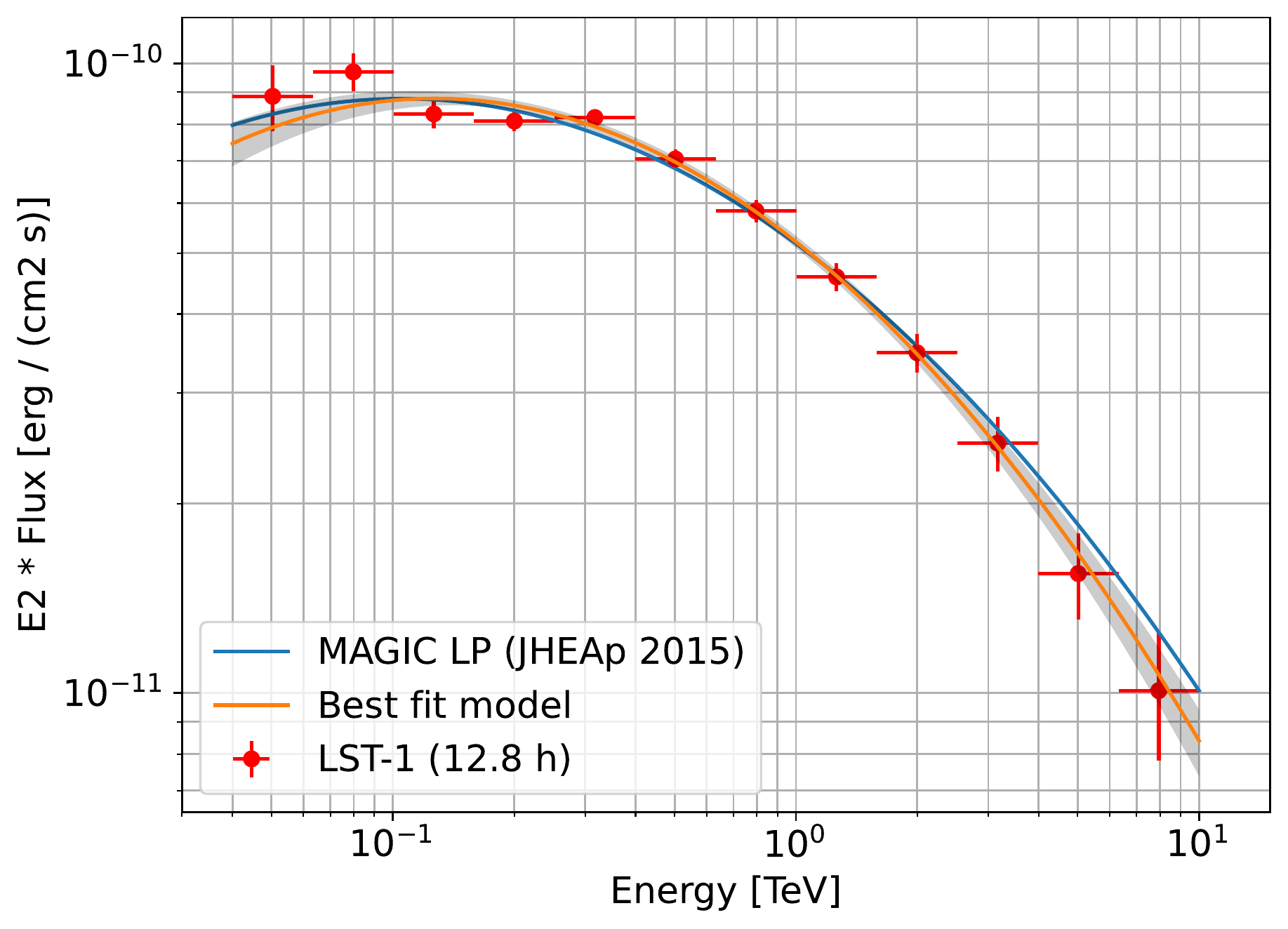}
\vspace{-0.3cm}
\caption{Left: Improvement of achieved flux sensitivity of LST1 over the last 2 years. Right: Spectral energy distribution of the Crab nebula as measured by LST1.}
\label{fig:sensitivity}
\end{center}
\end{figure}


The commissioning phase 3 is still ongoing. The amount of time the telescope has been observing gamma rays
since January 2020 reached 250 hrs in May 2021. We expect to intensify the amount of data taken and processed towards the end of 2021, targeting a value of about 100 hrs per month, which is a typical value for an operating imaging atmospheric Cherenkov telescope.

During the commissioning phase several lessons have been learned, which triggered some minor modifications in LST1 and subsequent LSTs. Among them are:
\begin{itemize}
    \item The automatic parking procedure needs fine tuning, in particular in elevation because the telescope is fixed in two points which are fixed on independent foundations: the central platform and the camera access tower.
    \item The Azimuth Locking System needs a protection against falling ice.
    \item The central pin area needs a cover to prevent accumulation of ice.
    \item The design of CMOS camera housing for the AMC system is not water tight.
    \item The camera calibration box with the laser needs to be improved to be able to monitor the light intensity of the laser with higher precision and to monitor the environmental conditions inside the box during the day.
\end{itemize}

The deficiencies above do not impose any large redesign and allow continuing telescope commissioning and fine tuning. The achieved low level performance of LST1 as well as physics performance so far is close to the one expected. The telescope performance continues to improve thanks to a work of a dedicated analysis team, see Figure~\ref{fig:sensitivity} for the example of the flux sensitivity (left) and the spectral energy distribution of the Crab nebula (right). It is an important achievement that the sensitivity of a single dish LST1 is close to the one obtained with a stereoscopic system of two MAGIC telescopes. Still, some work needs to be done in understanding the effect of light throughput of the telescope and the Cherenkov light image contamination by light produced by dim stars.

The LST camera calibration and commissioning are described in \citep{calibration, cameracommissioning}, 
and the monitoring of the telescope pointing in \citep{pointing}. 
Results of the first joint observations between LST1 and MAGIC can be found in \cite{xcalibration}. The work on the real time analysis using LST data is reported in \citep{gammalearn}, while new analysis methods are explored in
\citep{likelihood}. Development status of a novel LST camera based on Silicon photomultipliers can be found in \citep{sipm}. 

\begin{figure}
\begin{center}
\includegraphics[width=0.325\textwidth]{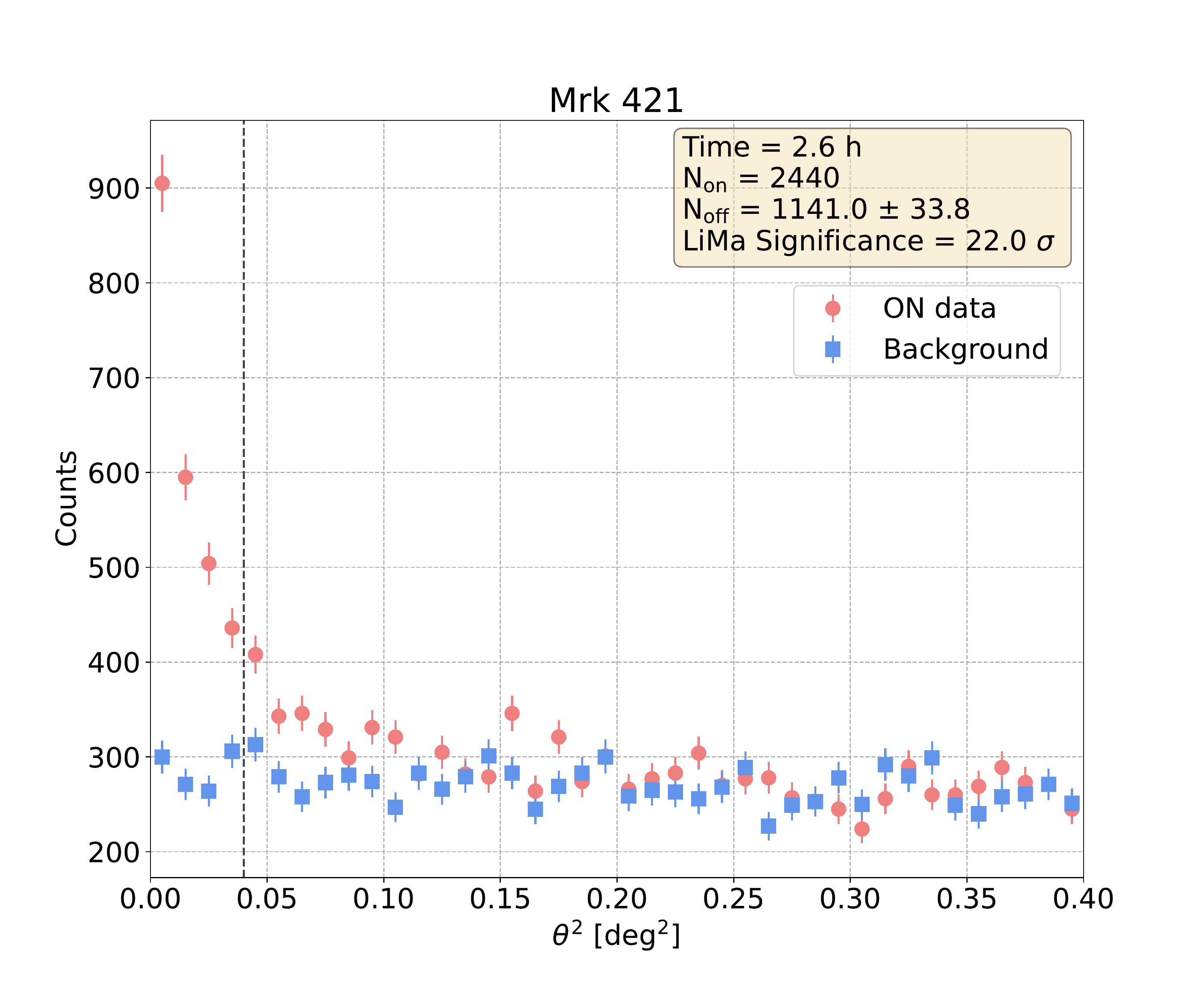}
\includegraphics[width=0.325\textwidth]{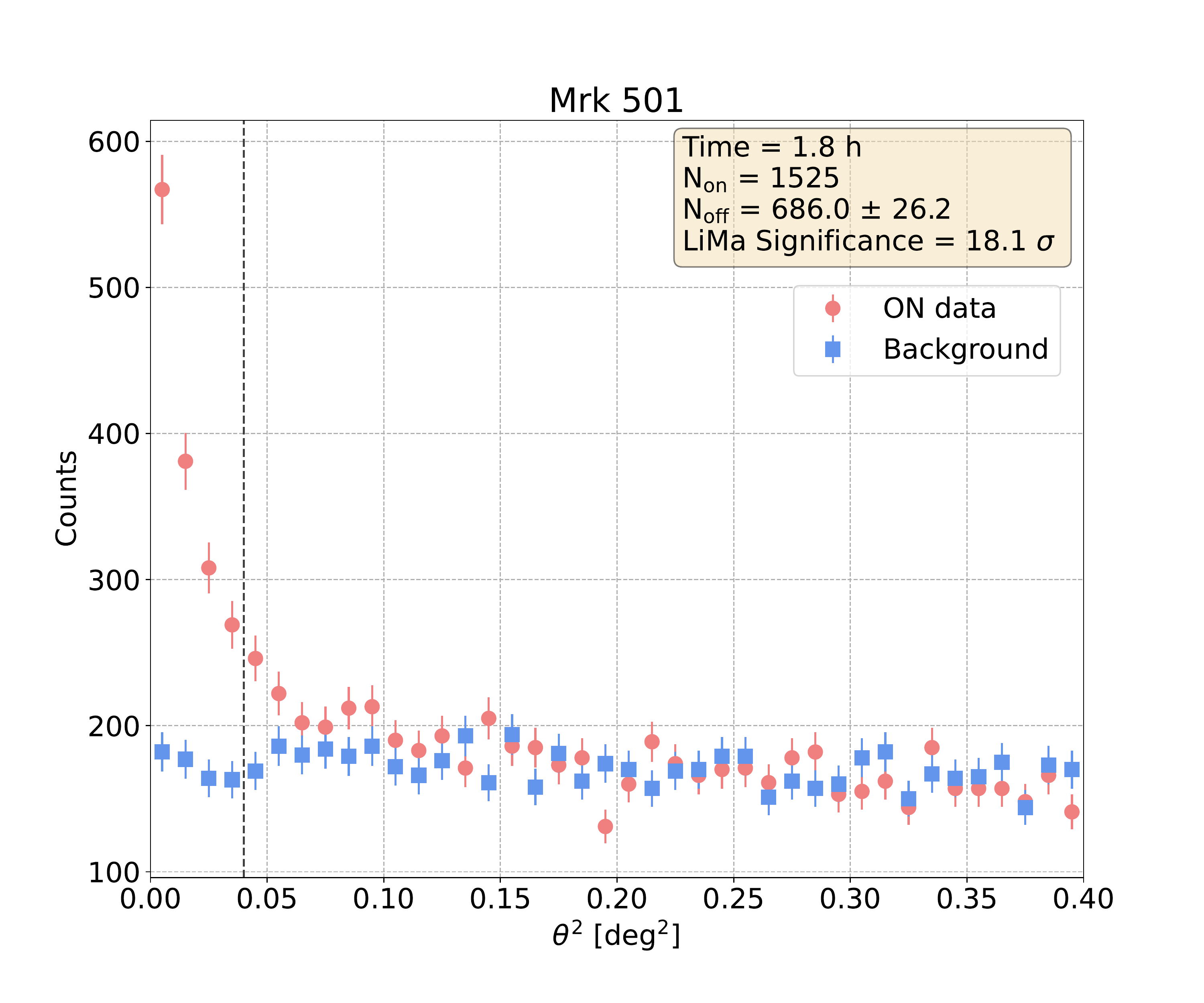}
\includegraphics[width=0.325\textwidth]{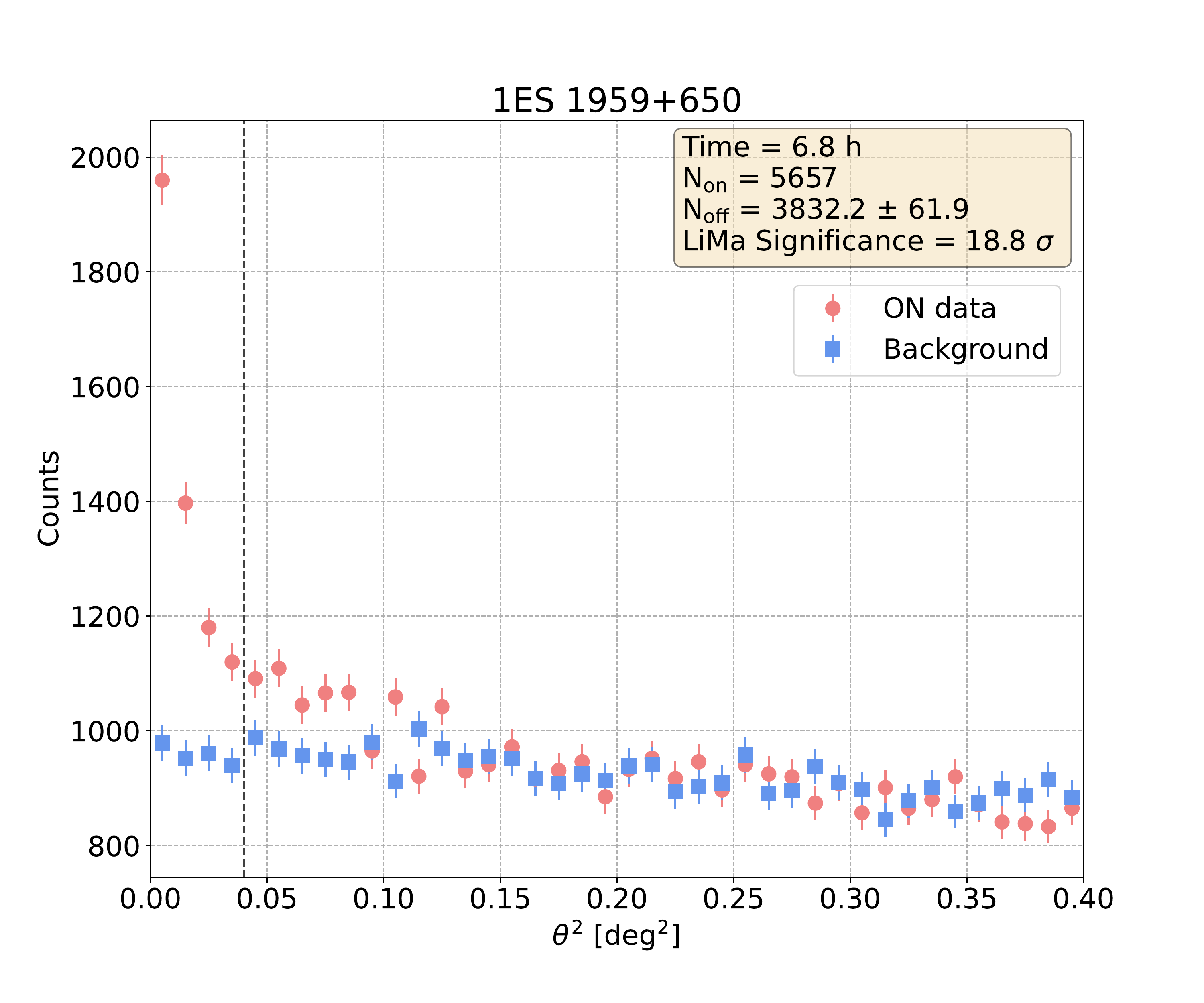}
\vspace{-0.3cm}
\caption{Detection plots of Mrk 421, Mrk 501 and 1ES 1959+650 with LST1.}
\label{fig:agns}
\end{center}
\end{figure}

As a part of the commissioning, LST1 is taking technical runs in the direction of known emitters of VHE $\gamma$ rays. 
The standard candles of the $\gamma$-ray astrophysics, the Crab nebula and the Crab pulsar, are observed on a regular basis to improve the LST1 performance and monitor the stability of the detector, see \citep{lst_performance} for details. 
Other good targets for LST1 are known VHE $\gamma$-ray blazars. As of June 2021, the following blazars have been detected with LST1: Markarian 421 (redshift z=0.031), Markarian\,501 (z=0.034), 1ES\,1959+650 (z=0.049), see Figure~\ref{fig:agns}, as well as PG1553+113 (redshift around z=0.5) and 1ES\,0647+250 (redshift is uncertain z$>$0.29).
The detected $\gamma$-ray signals prove crucial characteristics of the LSTs such as the low energy threshold and sensitivity to $\gamma$-rays below 50 GeV.




\section{LST2-4 status}
\label{sec:lst24}

\begin{figure}
\begin{center}
\includegraphics[width=0.95\textwidth]{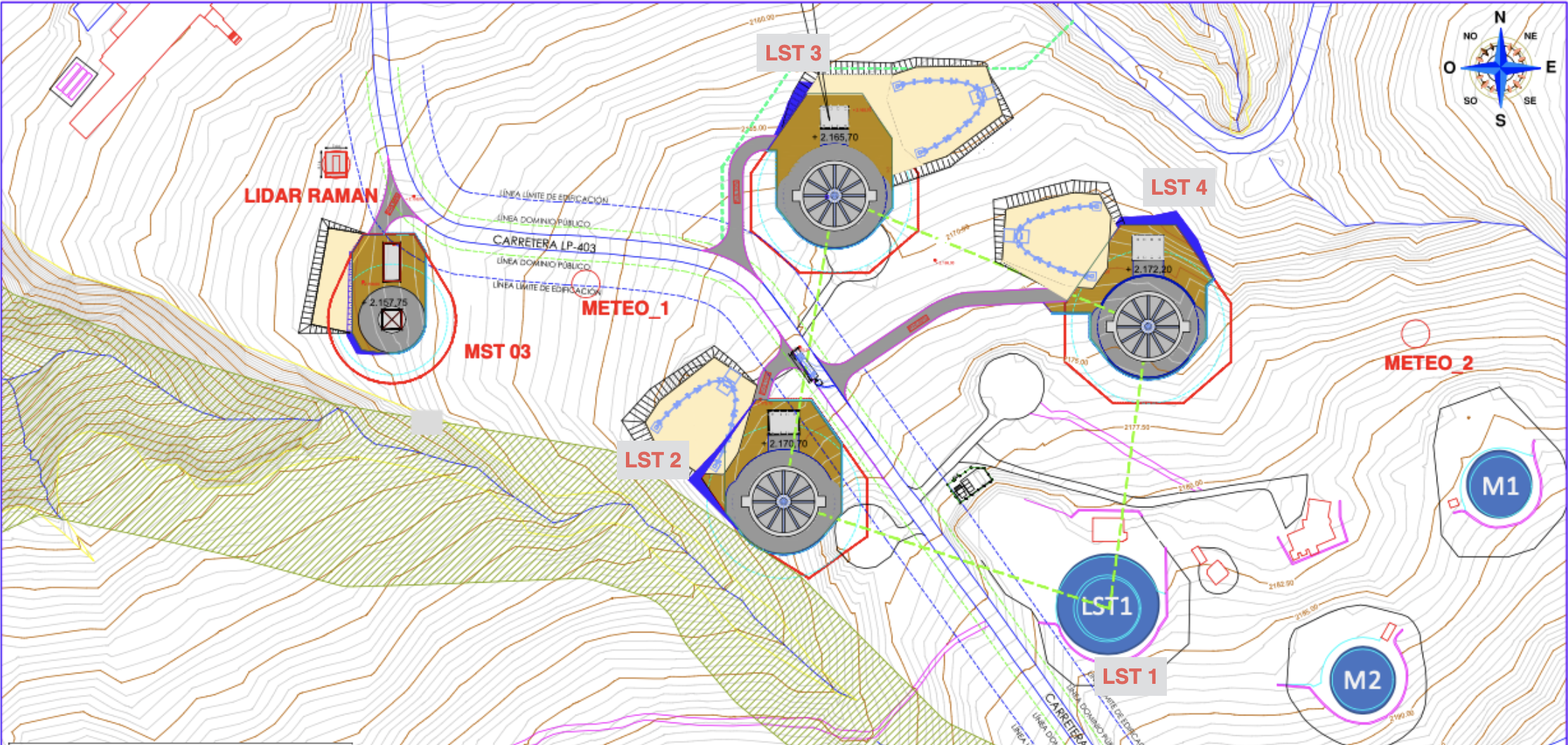}
\vspace{-0.3cm}
\caption{Locations of the LSTs in the CTA North.}
\label{fig:lst24}
\end{center}
\end{figure}

The preparation for the construction of LST2-4 at CTA North is under way. The complete funding for the construction and commissioning has been secured. The locations of the telescopes can be seen in Figure~\ref{fig:lst24}.
It is expected that the construction license will be obtained in Summer 2021 and the ground breaking will be in Autumn 2021.
In parallel, the production of telescope pieces and parts as well as their quality control is ongoing under contracts with the member institutes and coordination by the LST management. If this schedule can be kept, the LST2-4 will be erected until end of 2024, so that first scientific data from the northern LST array could be expected in 2025.

\section{Conclusion and Outlook}
\label{sec:conclusion}

We presented the status of the CTA LST project. The LST1 is finishing its commissioning phase and the performance is well within the expectations so that the construction of LST2-4 can start in La Palma in 2021.
First scientific results start to come using LST1 even though the expected sensitivity of a single telescope is lower than the one of the current IACTs. With the addition of further LSTs and several MSTs, CTA North will fulfill high expectations of the scientific community. 


\section*{Acknowledgements}
We gratefully acknowledge financial support from the agencies and organizations listed here: \url{http://www.cta-observatory.org/consortium_acknowledgments}

\clearpage
\section*{Full Authors List: \Coll\ collaboration}

\scriptsize
\noindent
H. Abe$^{1}$,
A. Aguasca$^{2}$,
I. Agudo$^{3}$,
L. A. Antonelli$^{4}$,
C. Aramo$^{5}$,
T.  Armstrong$^{6}$,
M.  Artero$^{7}$,
K. Asano$^{1}$,
H. Ashkar$^{8}$,
P. Aubert$^{9}$
% ...
\\

\noindent
$^{1}$Institute for Cosmic Ray Research, University of Tokyo.
$^{2}$Departament de Física Quàntica i Astrofísica, Institut de Ciències del Cosmos, Universitat de Barcelona, IEEC-UB.
$^{3}$Instituto de Astrofísica de Andalucía-CSIC.
$^{4}$INAF - Osservatorio Astronomico di Roma.
$^{5}$INFN Sezione di Napoli.
$^{6}$Aix Marseille Univ, CNRS/IN2P3, CPPM.
$^{7}$Institut de Fisica d'Altes Energies (IFAE), The Barcelona Institute of Science and Technology.
$^{8}$IRFU, CEA, Université Paris-Saclay.
$^{9}$LAPP, Univ. Grenoble Alpes, Univ. Savoie Mont Blanc, CNRS-IN2P3, Annecy.
% ...

%
%
%

\end{document}